\documentclass{article}
\usepackage{amsmath}
\usepackage{amssymb}

\begin{document}

\title{On the Security of Liaw et al.'s Scheme}

\author{Amit K Awasthi\\Department of Mathematics,\\Pranveer Singh Institute of Technology\\Kanpur-208020, UP, India.\\
Email: {awasthi@psit.in}}
\date{}
\maketitle

\begin{abstract}
Recently, Liaw et al. proposed a remote user authentication scheme
using smartcards. They claimed a number of features of their scheme, e.g. a dictionary of verification tables is not required to authenticate users; users can choose their password freely; mutual authentication is provided between the user and the remote system; the communication cost and the computational cost are very low; users can update their password after the registration phase;  a session key agreed by the user and the remote system is generated in every session; and the nonce-based scheme which does not require a timestamp (to solve the serious time synchronization problem) etc.

In this paper We show that Liaw et al.'s scheme does not stand with various security requirements and is completely insecure. \vspace{2 mm}\\
Keywords: Authentication, Smartcards, Remote system, Attack.
\end{abstract}

\section{Introduction}
In insecure communication network a remote user authentication is
a tool to authenticate remote users. Remote user authentication is a process by which a remote system
gains access to the remote resources. 

In 1981,Lamport \cite{lam81} proposed a password based remote user authentication scheme using password tables to verify the
remote user over insecure communication channel. That scheme was not fulfilling the security requirements in current senario. Since the Lamport's scheme , several remote user authentication schemes and improvements \cite{awa03}, \cite{hwa00a}, \cite{hwa00}, \cite{lia06}, \cite{yoo05} have been proposed with and without smart cards.  Some of these schemes are also discussed in a survey \cite{tsa06}. Recently, Liaw et al. \cite{lia06} proposed a remote user authentication scheme using smart cards. Their scheme has claimed a number of features , e.g. a dictionary of verification tables is not required to authenticate users; users can choose their password freely; mutual authentication is provided between the user and the remote system; the communication cost and the computational cost are very low; users can update their password after the registration phase;  a session key agreed by the user and the remote system is generated in every session; and the nonce-based scheme which does not require a timestamp (to solve the serious time synchronization problem) etc. In this paper We show that Liaw et al.'s scheme has many security holes and is completely insecure. 
\section{The Liaw et al.'s scheme} The scheme consists of five phases: registration, login,
verification, session and password change.

\subsection{Registration phase}
A new user $U_i$ submits identity $ID_i$ and password $PW_i$ to the remote system for registration.
The remote system computes $U_i$'s secret information $v_i =
h(ID_i, x)$ and $e_i = v_i \oplus PW_i$, where $x$ is a secret key
maintained by the remote system and $h(\cdot)$ is a secure one-way
hash function. Then the remote system writes $h(\cdot)$ and $e_i$
into the memory of a smart card and issues the card to $U_i$.

\subsection{Login phase} When $U_i$ wants to log into the remote
system, he/she inserts the smart card into the terminal and enters
$ID_i$ and $PW_i$. The smart card then performs the following
operations:
\newcounter{1}
\begin{list}{L\arabic{1}.}
{\usecounter{1}} \item Generate a random nonce $N_i$ and compute
$C_i = h(e_i \oplus PW_i, N_i)$. \item Send the login message $<
ID_i, C_i, N_i >$ to the remote system.
\end{list}

\subsection{Verification phase} To check the authenticity of $< ID_i,
C_i, N_i >$, the remote system checks the validity of $ID_i$. If
$ID_i$ is valid, computes $v_i^{\prime} = h(ID_i, x)$ and checks
whether $C_i = h(v_i^{\prime}, N_i)$. Then generates a random
nonce $N_s$, encrypts the message $M = E_{v_i^{\prime}}(N_i, N_s)$
and sends it back to the card.\\
The smart card decrypts the message $D_{e_i \oplus PW_i}(M)$ and
gets $(N_i^{\prime}, N_s^{\prime})$. Then verifies whether
$N_i^{\prime} = N_i$ and $N_s^{\prime} = N_s$. If
these checks hold valid, the mutual authentication is done.

\subsection{Session phase} This phase involves two public parameters
$q$ and $\alpha$ where $q$ is a large prime number and $\alpha$ is
a primitive element mod $q$. The phase works as follows:
\newcounter{5}
\begin{list}{S\arabic{5}.}
{\usecounter{5}}\item The remote system computes $S_i =
\alpha^{N_s}$ mod $q$ and sends $S_i$ to the smart card. The smart
card computes $W_i = \alpha^{N_i}$ mod $q$ and sends $W_i$ to the
remote system. \item The remote system computes $K_s =
(W_i)^{N_s}$ mod $q$ and, the smart card computes $K_u =
(S_i)^{N_i}$ mod $q$. It is easy to see that $K_s = K_u$. Then,
the card and the remote system exchange the data using the session
key and $e_i$.
\end{list}

\subsection{Password change phase} With this phase $U_i$ can change
his/her $PW_i$ by the following steps:
\newcounter{6}
\begin{list}{S\arabic{6}.}
{\usecounter{6}}\item Calculate $e_i^{\prime} = e_i \oplus PW_i
\oplus PW_i^{\prime}$. \item Update $e_i$ on the memory of smart
card to set $e_i^{\prime}$.
\end{list}

\section{Security Weaknesses}
\begin{list} {\arabic{1}.}{\usecounter{1}}
\item In registration phase user $U_i$ submits its identity $ID_i$ and Password $PW_i$ to the remote system. Medium of communication is not described. Is it secure or insecure. In real problems, user normally uses insecure channel. In such case password $PW_i$ is reveled to adversary $\mathcal{A}$ in between. 
\item In Login phase, when user $U_i$ keys his identity $ID_i$ and Password $PW_i$, smartcard computes a login message $<ID_i, C_i, N_i>$, Where $N_i$ is a random nonce and $C_i = h(e_i \oplus PW_i, N_i)$. This login message travels through insecure public channels. The adversary $\mathcal{A}$ can intercepts the valid login request $< ID_i, C_i, N_i>$. 

Now, with this infomation, advesary $\mathcal{A}$ can play replay attack. He sends $< ID_i, C_i, N_i>$ to the remote system at any time, as a login request . To validate $< ID_i,C_i, N_i >$, the remote system does the following:

\newcounter{11}
\begin{list}{-}
{\usecounter{11}} 
\item Checks the validity of $ID_i$. 
\item Computes $v_i^{\prime} = h(ID_i, x)$ and checks whether $C_i = h(v_i^{\prime}, N_i)$. Note this point, there is no check at the server side
which prevents the reuse of nonce $N_i$, which was already used in some previous login. Thus the server is unable to decide whether the $C_i$ is coming from a legitimate user or from an adversary. It is obvious that system authenticates the login request. 
\item The remote system generates a nonce $N_s^*$ and encrypts the message $M =E_{v_i^{\prime}}(N_i, N_s^*)$, then sends $<M>$ back to the
communicating party (that is advesary $\mathcal{A}$ here and is impersonating the legtimate user). 
\item Now, $\mathcal{A}$ will just reply 'OK' and will enjoy the access to the remote system. Therefore, ultimately the concept of mutual authentication fails on both side.
\end{list}

\item In above paragraph, adversary $\mathcal{A}$, has knowledge of login request $< ID_i,C_i, N_i >$. If he is able to access user's smartcard any how, he can recover the infomation $e_i$, which is stored on smartcard. Now having knowledge of $C_i$ and $e_i$, the adversary can perform offline attack, as he knows Three variables of the equation $C_i = h(e_i \oplus PW_i, N_i)$. He can hit and try various combination of passwords.

\item Session phase of Liaw et al.'s scheme is suffered from man-in-the-middle attack while the user and server are establishing common session key. It works as -
\newcounter{15}
\begin{list}{\arabic{15}.}
{\usecounter{15}}\item The remote system computes $x_S =
\alpha^{N_s^*} \mod q$ and communicates $x_S$. The adversary $\mathcal{A}$ computes $x_{\mathcal{A}} =\alpha^{N_i} \mod q$ and sends $x_{\mathcal{A}}$ to the remote system. 
\item
The remote system computes $K_s = (x_{\mathcal{A}})^{N_s^*}$ mod $q$ and
$\mathcal{A}$ computes $K_a = (x_S)^{N_i}$ mod $q$. It is easy to see that $K_s = K_a$. Now with the help of other public parameters adversary can communicate with server in encrypted way.
\end{list}
\end{list}
\section{Conclusion}
In this paper, we have shown various security holes of the Liaw et al.'s scheme.

\end{document}